\documentstyle[12pt]{article}
\begin{document}
\title{ A reformulation of the Ponzano-Regge quantum gravity model
in terms of surfaces}
\author{Junichi Iwasaki\\
Physics Department, University of Pittsburgh\\
 Pittsburgh, PA 15260, USA.\\
E-mail: iwasaki@phyast.pitt.edu}
\date{Jan. 11, 1994; revised on Sep. 29, 1994}

\maketitle

\begin{abstract}

We reformulate the Ponzano-Regge quantum gravity model
in terms of surfaces on a 3-dimensional simplex lattice.
This formulation
(1) has a clear relation to the loop representation of the canonical
quantum general relativity in 3-dimensions,
(2) may have a 4-dimensional analogue, in contrast to
the 6-j symbolic formalism of the Ponzano-Regge model, and
(3) is purely a theory of surfaces, in the sense that
it does not include any field variables;
hence it is coordinate-free on the surface
and background-free in spacetime.
We discuss implications and applications of this formulation.

\end{abstract}

\section{Introduction}
\label{sec:intro}

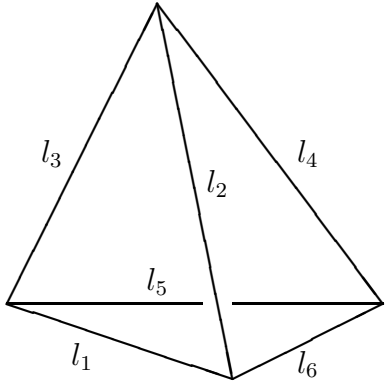
\begin{figure}
\unitlength=1.0mm
%\linethickness{0.6pt}
\begin{picture}(90.00,55.00)(-40.00,-15.00)
\thicklines
\put(0.00,0.00){\line(1,2){20.00}}
\put(0.00,0.00){\line(3,-1){30.00}}
\put(20.00,40.00){\line(1,-5){10.0}}
\put(20.00,40.00){\line(3,-4){30.0}}
\put(30.00,-10.00){\line(2,1){20.0}}
\put(0.00,0.00){\line(1,0){26.0}}
\put(50.0,0.00){\line(-1,0){20.0}}
\put(10.00,-7.0){\makebox(0,0)[cc]{$l_1$}}
\put(28.0,16.0){\makebox(0,0)[cc]{$l_2$}}
\put(20.0,3.0){\makebox(0,0)[cc]{$l_5$}}
\put(6.00,20.0){\makebox(0,0)[cc]{$l_3$}}
\put(40.00,20.0){\makebox(0,0)[cc]{$l_4$}}
\put(40.00,-8.0){\makebox(0,0)[cc]{$l_6$}}
\end{picture}
\caption{The tetrahedron with edge lengths $l_i\ (i=1,2,\cdots,6)$.}
\label{fig:tetra}
\end{figure}

The Ponzano-Regge quantum gravity model has been a mystery
since the time of its construction \cite{ponzano}.
Ponzano and Regge found an unexpected relation between
the 6-j symbol and Regge's discrete version of
the action functional of
(Euclidean) general relativity  in 3-dimensions.
The description of the relation follows.
Consider a 3-dimensional simplex manifold, which consists of
tetrahedra whose faces are attached to one another.
Pick a tetrahedron and denote the lengths of its edges by
$l_i\ (i=1,2,\cdots,6)$ as shown in Fig.\ref{fig:tetra}.
The Regge action for the tetrahedron is
\begin{equation}
S_{Regge}=\sum_{i=1}^6 l_i\theta_i,
\end{equation}
where $\theta_i$ is the angle between the outward normals
of two faces sharing the $i$-th edge.
If we assign half-integer values $j_i\ (i=1,2,\cdots,6)$
to the edges such that $l_i=j_i+{1\over2}$, then the 6-j
symbol $\left\{\matrix{j_1&j_2&j_3\cr j_4&j_5&j_6\cr}\right\}$
 approximates the cosine of the Regge action
in the large $j$ limit.
The positive frequency part of the cosine function is
the integrand of the partition function of quantum (Euclidean)
general relativity
(up to some subtleties concerning the imaginary unit
$i\equiv\sqrt{-1}$) in the pathintegral formalism.
Based on this observation, Ponzano and Regge constructed a
``regularized" partition function of quantum gravity
on the simplex manifold
in terms of 6-j symbols.

Recently  by constructing
 the same model in terms of a quantum group,
Turaev and Viro \cite{turaev} proved that the partition
function of this model is independent of the choice of
the tetrahedral decomposition of 3-manifold.
{}From this fact it could be said that the partition function
of this model is invariant under the renormalization group
transformation and depends  only on the topology of the manifold.
This could be an indication that the model is related to some
continuum quantum theory.

By constructing a Hilbert space of
state functionals on a 2-space
formed and triangulated by the faces of the tetrahedra
in the simplex manifold,
Ooguri \cite{ooguri} showed that the model is isomorphic
to Witten's canonical formulation of $ISO(3)$ Chern-Simons
theory \cite{witten}, which is known to be equivalent to
the Ashtekar formulation of general relativity in 3-dimensions
\cite{2+1,abhay}.
(In Ooguri's construction the relevant group is
$SO(3)$ rather than $SO(2,1)$.)

Afterward, Rovelli \cite{rovelli} pointed out
that Ooguri's construction
is closely related to the loop representation of canonical
quantum general relativity \cite{loop}
and constructed the basis of the loop representation
on the 2-space from
the basis of Ooguri's state space.
In other words, what Rovelli showed was a way of constructing
a loop transformation from the connection representation to
the loop representation through Ooguri's state space.
Then, Rovelli proposed an idea of constructing the inner product
of the loop representation using the transition amplitude
consisting of 6-j symbols.

However, the 6-j symbolic form of the transition amplitude
is not directly related to the basis of the loop representation
but to the basis of Ooguri's state space.
Therefore, it is desired to find some reconstruction of
the transition amplitude, if it exists, which is directly
related to the basis of the loop representation
in the same way that the 6-j symbolic construction is
directly related to the basis of Ooguri's state space.

In this paper, we demonstrate that such a reconstruction exists.
We reconstruct the partition function of the model
in terms of a sum over surfaces in such a way that
the intersections of these surfaces with any 2-space,
consisting of the faces of tetrahedra, are loops.
It turns out that these loops are the loops which are
associated with the basis of the loop representation
constructed from the basis of Ooguri's state space.

This space-time formulation in terms of surfaces are
equivalent to the space-time formulation in terms of
6-j symbols, namely the Ponzano-Regge model,
in the same sense that
the canonical loop representation is equivalent to
the representation on Ooguri's state space.
The construction of the canonical loop representation
from the space-time formulation in terms of surfaces
is parallel to the construction of the canonical representation
on Ooguri's state space from the space-time formulation
in terms of 6-j symbols.

Since the relation between the connection and loop representations,
namely the loop transformation, was already constructed by Rovelli
by making use of the basis of Ooguri's state space,
the aim of this paper is not to show any relation to
connection formulations but to show
the existence of a ``duality" between
the space-time formulation and the canonical representation
within the use of surfaces and loops without any variables
(i.e. connections).

Eventually we would like to make use of this reformulation
to rederive the inner product of
the (2+1) loop representation, in which loop states are
labelled by homotopy classes.
To do so, however, we have to regularize infinities present
in the partition function, which is not yet under control
in the language of surfaces.
Therefore, we leave these issues for future work
but suggest a possible strategy at the end of the paper.

In Sec.\ref{sec:surface} we describe our reformulation of
the model.
In Sec.\ref{sec:discuss} we discuss
implications and  applications of our formulation.

\section{The construction of surfaces from 6-j symbols}
\label{sec:surface}

The partition function of the Ponzano-Regge model
on a 3d simplex manifold $M$
is defined by
\begin{equation}
Z_M:=\lim_{L\to\infty}\sum_{j\le L}\prod_{vertices}
\Lambda^{-1}(L)\prod_{edges}(2j+1)
\prod_{\scriptstyle tetra-\atop\scriptstyle hedra}
(-1)^{\sum_ij_i}\left\{\matrix{j_1 & j_2 & j_3\cr
j_4 & j_5 & j_6} \right\}.
\label{eq:part}
\end{equation}
Here $\Lambda(L)$ is introduced to regulate infinities
and is defined by
\begin{equation}
\Lambda(L):=\sum_{k=0,{1\over2},1,\cdots,L}(2k+1)^2,
\label{eq:lambda}
\end{equation}
which behaves as
$\Lambda(L)\sim{4\over3}L^3$ as $L\to\infty$.
We replace the 6-j symbol in Eq.(\ref{eq:part}) by some quantities
which are associated with surfaces
(more precisely, with perimeters of surfaces).
To do so, we make four observations below.

The first observation is that the 6-j symbol can be written
in terms of 3-j symbols (up to the factor of $2c+1$)\cite{6-j} as
\begin{eqnarray}
&\left\{\matrix{a & b & e\cr d & c & f}\right\}
&= (-1)^{-e-f}
\sum_{\scriptstyle\alpha,\beta,\gamma
\atop\scriptstyle\delta,\varepsilon,\phi}
(-1)^{-\alpha-\delta}\nonumber\\
&&\times
\left(\matrix{a & b & e\cr \alpha & \beta & -\varepsilon}\right)
\left(\matrix{d & c & e\cr \delta & \gamma & \varepsilon}\right)
\left(\matrix{d & b & f\cr \delta & \beta & -\phi}\right)
\left(\matrix{a & c & f\cr \alpha & \gamma & \phi}\right).
\label{eq:6-j}
\end{eqnarray}

\begin{figure}
\unitlength=1.0mm
%\linethickness{0.6pt}
\begin{picture}(120.00,30.00)(-10.00,-10.00)
{\thicklines
\put(0.00,0.00){\line(1,0){40.00}}
\put(10.00,20.00){\line(1,0){40.00}}
\put(0.00,0.00){\line(1,2){10.0}}
\put(10.00,20.00){\line(1,-2){10.0}}
\put(20.00,0.00){\line(1,2){10.0}}
\put(30.00,20.00){\line(1,-2){10.0}}
\put(40.0,0.00){\line(1,2){10.0}}
}
\put(18.00,12.0){\makebox(0,0)[cc]{$j_1$}}
\put(3.0,12.0){\makebox(0,0)[cc]{$j_2$}}
\put(47.0,8.0){\makebox(0,0)[cc]{$j_2$}}
\put(28.0,8.0){\makebox(0,0)[cc]{$j_5$}}
\put(10.00,-3.0){\makebox(0,0)[cc]{$j_3$}}
\put(30.00,-3.0){\makebox(0,0)[cc]{$j_3$}}
\put(38.00,12.0){\makebox(0,0)[cc]{$j_4$}}
\put(20.00,23.0){\makebox(0,0)[cc]{$j_6$}}
\put(40.00,23.0){\makebox(0,0)[cc]{$j_6$}}

\put(70.00,0.00){\line(1,0){40.00}}
\put(80.00,20.00){\line(1,0){40.00}}
\put(70.00,0.00){\line(1,2){10.0}}
\put(80.00,20.00){\line(1,-2){10.0}}
\put(90.00,0.00){\line(1,2){10.0}}
\put(100.00,20.00){\line(1,-2){10.0}}
\put(110.0,0.00){\line(1,2){10.0}}

{\thicklines
\put(80.0,5.0){\line(0,-1){10.0}}
\put(80.0,5.0){\line(-1,1){10.0}}
\put(80.0,5.0){\line(1,1){10.0}}
\put(90.0,15.0){\line(0,1){10.0}}
\put(90.0,15.0){\line(1,-1){10.0}}
\put(100.0,5.0){\line(0,-1){10.0}}
\put(100.0,5.0){\line(1,1){10.0}}
\put(110.0,15.0){\line(0,1){10.0}}
\put(110.0,15.0){\line(1,-1){10.0}}
}
\put(90.00,10.0){\makebox(0,0)[cc]{$j_1$}}
\put(70.0,10.0){\makebox(0,0)[cc]{$j_2$}}
\put(120.0,10.0){\makebox(0,0)[cc]{$j_2$}}
\put(100.0,11.0){\makebox(0,0)[cc]{$j_5$}}
\put(85.00,-3.0){\makebox(0,0)[cc]{$j_3$}}
\put(105.00,-3.0){\makebox(0,0)[cc]{$j_3$}}
\put(110.00,10.0){\makebox(0,0)[cc]{$j_4$}}
\put(95.00,23.0){\makebox(0,0)[cc]{$j_6$}}
\put(115.00,23.0){\makebox(0,0)[cc]{$j_6$}}
\end{picture}
\caption{The 2d lattice (left) and its dual lattice
(right) on the faces of a single tetrahedron.}
\label{fig:latt}
\end{figure}
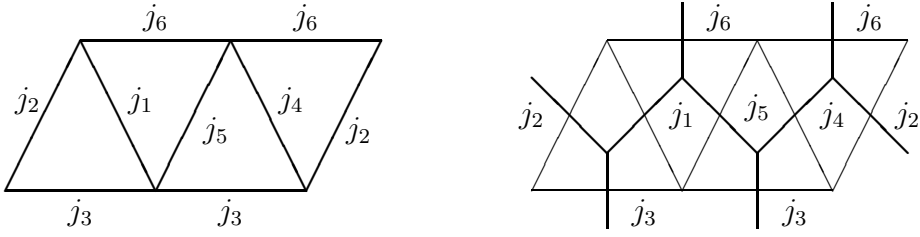

The second obsevation is that by defining matrices
\begin{eqnarray}
&&U^{\ m}_{(j)\ n}:=(-1)^{j-2m}\delta^m_{-n}, \nonumber\\
&&V^{\ m}_{(j)\ n}:=(-1)^{j-m}\delta^m_{-n},\nonumber\\
&&W^{\ m}_{(j)\ n}:=(-1)^{m}\delta^m_{n},
\label{eq:mat}
\end{eqnarray}
where $j$ is a half-integer and $m$ and $n$ take half-integer
values between $-j$ and $j$ (like the angular momentum),
Eq.(\ref{eq:6-j}) can be rewritten as
\begin{eqnarray}
&(-1)^P\left\{\matrix{a & b & e\cr d & c & f}\right\}
&=
%\sum_{\scriptstyle\alpha,\beta,\gamma,\delta,\varepsilon,\phi
%\atop\scriptstyle\alpha',\beta',\gamma',
%\delta',\varepsilon',\phi'}
\sum
(-1)^{\alpha+\beta+\gamma+\delta+\varepsilon+\phi}\nonumber\\
&&\times
\left(\matrix{a & b & e\cr \alpha' & \beta' & \varepsilon'}\right)
\left(\matrix{d & c & e\cr \delta' & \gamma' & \varepsilon}\right)
\left(\matrix{d & b & f\cr \delta & \beta & \phi'}\right)
\left(\matrix{a & c & f\cr \alpha & \gamma & \phi}\right)\nonumber\\
&&\times U^{\ \alpha'}_{(a)\ \alpha''}V^{\ \beta'}_{(b)\ \beta''}
V^{\ \gamma'}_{(c)\ \gamma''}U^{\ \delta'}_{(d)\ \delta''}
W^{\ \varepsilon'}_{(e)\ \varepsilon''}W^{\ \phi'}_{(f)\ \phi''}
\nonumber\\
&&\times\delta^{-\alpha'',\alpha}\delta^{-\beta'',\beta}
\delta^{-\gamma'',\gamma}\delta^{-\delta'',\delta}
\delta^{-\varepsilon'',\varepsilon}\delta^{-\phi'',\phi}
\label{eq:triv}
\end{eqnarray}
with $P:=a+b+c+d+e+f$,
where the sum is taken with respect to all Greek letters.
Notice that the matrices $U_{(j)}$, $V_{(j)}$ and $W_{(j)}$
can be seen as the spin-$j$ representation matrices of $SU(2)$
elements. Indeed, when $j={1\over2}$ they are $-i$ times
the Pauli matrices $-i\sigma_1$, $-i\sigma_2$ and $-i\sigma_3$
respectively.
Consider the faces of the tetrahedron associated
with the 6-j symbol as a {\it 2-dimensional simplex lattice} with
the topology of the sphere.
For later use, pictures of the lattice and its dual lattice are
shown in Fig.\ref{fig:latt}.
Then the 6-j symbol can be seen
as an $SU(2)$ gauge invariant function defined by Eq.(\ref{eq:triv})
on the 2d dual lattice.
Here $j$'s associated with the six edges
are assigned to the corresponding dual lines and the matrices are
understood as if they were parallel transports along  the dual lines.
The matrices are assigned such that the three matrices,
$U_{(j)}$, $V_{(j)}$ and $W_{(j)}$ with different $j$'s in general,
meet together at every triangle.
Actually, these matrices are not parallel transports since we do not
have any connection variables in this formulation.
In this paper, we loosely use the terms ``parallel transports" and
``holonomies" to refer to these matrices and matrices contracted
from them along loops respectively.
Also we use terms sites, lines, triangles on 2d lattices,
instead of vertices, edges, faces, (tetrahedra), which are used in
a 3-dimensional sense.

The third observation is that
 Eq.(\ref{eq:triv}) can be written as a linear combination
of products of traces of ``holonomies" consisting of
$U_{(1/2)}$, $V_{(1/2)}$ and/or $W_{(1/2)}$.
This is possible because each component of a spin-$j$ matrix can be
written as a sum of products of the components of the corresponding
spin-${1\over2}$ matrix.
To understand this process pictorially,
we consider the following manipulations on the 2d dual lattice.
Given a half-integer value $j$ for a dual line, which means that
we have a ``parallel transport" in spin-$j$ representation,
place $2j$  ``parallel transports" in spin-${1\over2}$
representation on the dual line.
Do the same procedure
for all the other dual lines on the lattice.
Contract all the spin-${1\over2}$ matrices on the dual lattice
in all the possible ways.
Each way of doing the contractions determines, in general,
a product of traces of spin-${1\over2}$ ``holonomies"
along a (multiple) loop.

\begin{figure}
\unitlength=1.0mm
%\linethickness{0.6pt}
\begin{picture}(120.00,60.00)(-10.00,-10.00)
%\thicklines
% 4 triangles in the bottom
\put(0.00,0.00){\line(1,0){20.00}}
\put(0.00,0.00){\line(1,2){10.0}}
\put(10.00,20.00){\line(1,-2){10.00}}
{\thicklines
\put(10.00,5.00){\line(-1,1){10.00}}
\put(10.00,5.00){\line(1,1){10.00}}
\put(10.00,7.00){\line(-1,1){9.00}}
\put(10.00,7.00){\line(1,1){9.00}}
}
\put(10.00,-5.0){\makebox(0,0)[cc]{(e)}}

\put(30.00,0.00){\line(1,0){20.00}}
\put(30.00,0.00){\line(1,2){10.0}}
\put(40.00,20.00){\line(1,-2){10.00}}
{\thicklines
\put(39.00,5.00){\line(-1,1){9.00}}
\put(39.00,5.00){\line(1,1){10.00}}
\put(41.00,5.00){\line(-1,1){10.00}}
\put(41.00,5.00){\line(1,1){9.00}}
}
\put(40.00,-5.0){\makebox(0,0)[cc]{(f)}}

\put(60.00,0.00){\line(1,0){20.00}}
\put(60.00,0.00){\line(1,2){10.0}}
\put(70.00,20.00){\line(1,-2){10.00}}
{\thicklines
\put(70.00,5.00){\line(-1,1){10.00}}
\put(70.00,5.00){\line(1,1){10.00}}
\put(70.00,7.00){\line(-1,1){9.00}}
\put(70.00,7.00){\line(1,1){9.00}}
\put(70.00,6.00){\circle*{4}}
}
\put(70.00,-5.0){\makebox(0,0)[cc]{(g)}}

\put(90.00,0.00){\line(1,0){20.00}}
\put(90.00,0.00){\line(1,2){10.0}}
\put(100.00,20.00){\line(1,-2){10.00}}
{\thicklines
\put(98.50,5.50){\line(-1,1){9.00}}
\put(99.50,6.50){\line(-1,1){9.00}}
\put(101.50,5.50){\line(1,1){9.00}}
\put(100.50,6.50){\line(1,1){9.00}}
\put(98.50,6.50){\oval(2,2)[br]}
\put(101.50,6.50){\oval(2,2)[bl]}
}
\put(100.00,-5.0){\makebox(0,0)[cc]{(h)}}

% 4 triangles in the top
\put(0.00,35.00){\line(1,0){20.00}}
\put(0.00,35.00){\line(1,2){10.0}}
\put(10.00,55.00){\line(1,-2){10.00}}
{\thicklines
\put(10.00,42.00){\line(-1,1){9.00}}
\put(10.00,42.00){\line(1,1){9.00}}
\put(10.00,40.00){\line(0,-1){10.00}}
\put(10.00,40.00){\line(1,1){10.00}}
}
\put(10.00,25.0){\makebox(0,0)[cc]{(a)}}

\put(30.00,35.00){\line(1,0){20.00}}
\put(30.00,35.00){\line(1,2){10.0}}
\put(40.00,55.00){\line(1,-2){10.00}}
{\thicklines
\put(41.00,40.00){\line(-1,1){10.00}}
\put(41.00,40.00){\line(1,1){10.00}}
\put(40.00,41.00){\line(0,-1){10.00}}
\put(40.00,41.00){\line(1,1){10.00}}
}
\put(40.00,25.0){\makebox(0,0)[cc]{(b)}}

\put(60.00,35.00){\line(1,0){20.00}}
\put(60.00,35.00){\line(1,2){10.0}}
\put(70.00,55.00){\line(1,-2){10.00}}
{\thicklines
\put(70.00,40.00){\line(-1,1){10.00}}
\put(70.00,42.00){\line(1,1){9.00}}
\put(70.00,40.00){\line(0,-1){10.00}}
\put(70.00,40.00){\line(1,1){10.00}}
\put(70.00,41.00){\circle*{4}}
}
\put(70.00,25.0){\makebox(0,0)[cc]{(c)}}

\put(90.00,35.00){\line(1,0){20.00}}
\put(90.00,35.00){\line(1,2){10.0}}
\put(100.00,55.00){\line(1,-2){10.00}}
{\thicklines
\put(100.00,40.00){\line(-1,1){9.00}}
\put(100.00,40.00){\line(0,-1){10.00}}
\put(101.50,40.50){\line(1,1){9.00}}
\put(100.50,41.50){\line(1,1){9.00}}
\put(101.50,41.50){\oval(2,2)[bl]}
}
\put(100.00,25.0){\makebox(0,0)[cc]{(d)}}

\put(25.00,10.0){\makebox(0,0)[cc]{$\equiv$}}
\put(55.00,10.0){\makebox(0,0)[cc]{$=:$}}
\put(85.00,10.0){\makebox(0,0)[cc]{$\ne$}}
\put(25.00,45.0){\makebox(0,0)[cc]{$\equiv$}}
\put(55.00,45.0){\makebox(0,0)[cc]{$=:$}}
\put(85.00,45.0){\makebox(0,0)[cc]{$\ne$}}
\end{picture}
\caption{The gluing of lines at the dual site, corresponding to
contractions of spin-${1\over2}$ parallel transports.}
\label{fig:spin}
\end{figure}
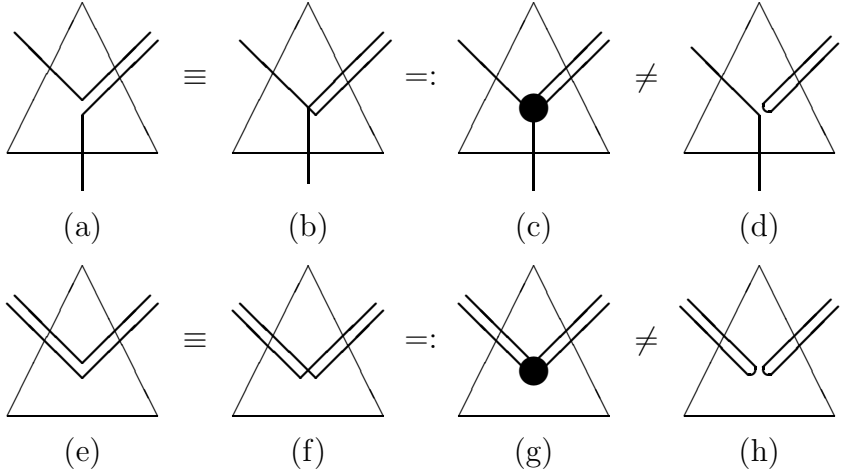

Notice that since the spin-${1\over2}$ ``parallel transports"
placed on the same dual line are identical and indistinguishable,
even though they are drawn as separate lines in the picture,
some different ways of doing the contractions give terms identical
in expression.
This fact can be used to define  equivalence classes of
(multiple) loops.
To express this fact pictorially, we identify in Fig.\ref{fig:spin}
 diagrams (a) and (b), or diagrams (e) and (f)
after  gluing lines (or contracting the corresponding
spin-${1\over2}$ matrices).
We denote them as (c) and (g) respectively.
Note that (d) and (h) are different
from (c) and (g) respectively.
Also note that since the freedom within
an equivalence class is due to the absence of a unique way
of doing the contractions of the matrices at a triangle,
it is related to the presence of
self-intersections of a (multiple) loop.
In this sense, this classification of loops is a ``regularization"
of otherwise uncontrollable intersecting loops.
In general, for a given configuration of $j$'s
there is more than one way of contracting the spin-${1\over2}$
matrices corresponding
to different equivalence classes of (multiple) loops.
 The 6-j symbol is now represented as a sum of terms,
each of which corresponds to an equivalence class of
(multiple) loops on the faces of the tetrahedron.

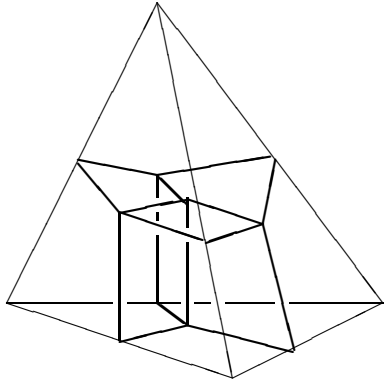
\begin{figure}
\unitlength=0.1mm
%\linethickness{0.6pt}
\begin{picture}(900.00,550.00)(-400.00,-150.00)
%\thicklines
% tetrahedron
\put(0.00,0.00){\line(1,2){200.00}}
\put(0.00,0.00){\line(3,-1){300.00}}
\put(200.00,400.00){\line(1,-5){100.0}}
\put(200.00,400.00){\line(3,-4){300.0}}
\put(300.00,-100.00){\line(2,1){200.0}}

%\put(0.00,0.00){\line(1,0){260.0}}
\put(0.00,0.00){\line(1,0){140.0}}
\put(160.00,0.00){\line(1,0){70.0}}
\put(250.00,0.00){\line(1,0){20.0}}
\put(290.00,0.00){\line(1,0){65.0}}
\put(500.0,0.00){\line(-1,0){125.0}}
%\put(500.0,0.00){\line(-1,0){200.0}}

%\put(100.00,-70.0){\makebox(0,0)[cc]{$j_1$}}
%\put(280.0,160.0){\makebox(0,0)[cc]{$j_2$}}
%\put(200.0,30.0){\makebox(0,0)[cc]{$j_5$}}
%\put(60.00,200.0){\makebox(0,0)[cc]{$j_3$}}
%\put(400.00,200.0){\makebox(0,0)[cc]{$j_4$}}
%\put(400.00,-80.0){\makebox(0,0)[cc]{$j_6$}}

% elementary surfaces
{\thicklines
\put(150.00,120.00){\line(0,-1){170.00}}
\put(150.00,120.00){\line(-5,6){55.00}}
\put(150.00,120.00){\line(3,-1){110.00}}
\put(150.00,120.00){\line(5,1){90.00}}

\put(200.00,170.00){\line(0,-1){30.00}}
\put(200.00,120.00){\line(0,-1){10.00}}
\put(200.00,90.00){\line(0,-1){90.00}}
%\put(200.00,170.00){\line(0,-1){170.00}}
\put(200.00,170.00){\line(-5,1){105.00}}
\put(200.00,170.00){\line(6,1){150.00}}
\put(200.00,170.00){\line(1,-1){40.00}}

\put(340.00,105.00){\line(1,-4){42.00}}
\put(340.00,105.00){\line(1,5){17.00}}
\put(340.00,105.00){\line(-3,1){100.00}}
\put(340.00,105.00){\line(-3,-1){75.00}}

\put(240.00,-30.00){\line(0,1){110.00}}
\put(240.00,100.00){\line(0,1){40.00}}
%\put(240.00,-30.00){\line(0,1){170.00}}
\put(240.00,-30.00){\line(-4,3){40.00}}
\put(240.00,-30.00){\line(-4,-1){90.00}}
\put(240.00,-30.00){\line(4,-1){140.00}}
}
\end{picture}
\caption{The six elementary surfaces in a tetrahedron.
Each one is a portion of a dual face.}
\label{fig:elm}
\end{figure}

The last observation is that there exists a unique surface
corresponding to a loop defined on the 2d dual lattice.
Let us consider the tetrahedron
as a {\it 3-dimensional simplex lattice}.
There are six dual faces corresponding to
the edges of the tetrahedron.
We define ``elementary surfaces", each of which is a portion of
a dual face and is inside the tetrahedron as shown in
Fig.\ref{fig:elm}.
Given a loop, we can always construct a surface consisting of
the elementary surfaces such that its perimeter is the loop.
Accordingly, given an equivalence class of loops,
we can define the corresponding equivalence class of surfaces.
In the rest of this paper, whenever we use the terms loops or surfaces,
we imply equivalence classes of (multiple) loops or surfaces respectively
unless otherwise explicitly stated.

The observations we made allow us to express
a 6-j symbol as a sum of terms, each of which is related to
a  surface bounded by a  loop
on the faces of the tetrahedron the 6-j symbol is associated with.
 Formally, the 6-j symbol associated with a tetrahedron
can be written, in fact, as
\begin{equation}
(-1)^{\sum_{i}j_i}
\left\{\matrix{j_1&j_2&j_3\cr j_4&j_5&j_6}\right\}=
\sum_{S\in E(j)}C^{\alpha(S)}_jT[\alpha(S)].
\label{eq:CT}
\end{equation}
Here $E(j)$ is the set of all the possible surfaces for given $j$'s,
$\alpha(S)$ is the perimeter of such a surface $S$, and
$T[\alpha(S)]$ is the product of traces of ``holonomies" along
the  loop $\alpha(S)$.
$T[\alpha(S)]$ contributes to the sign of the term
and its magnitude is always unity by construction.
The coefficient
$C^{\alpha(S)}_j$ is what we want to construct.
Therefore, the partition function can be written as
\begin{eqnarray}
&Z_M&=\lim_{L\to\infty}\sum_{j\le L}\prod_{vertices}
\Lambda^{-1}(L)\prod_{edges}(2j+1)
\prod_{\scriptstyle tetra-\atop\scriptstyle hedra}
\sum_{S\in E(j)}C_j^{\alpha(S)}T[\alpha(S)]
\label{eq:part1}\\
&&=\lim_{N\to\infty}\sum_{\bar S}\prod_{vertices}
\Lambda^{-1}(N)\prod_{edges}(n(\bar S)+1)\nonumber\\
&&\times
\prod_{\scriptstyle tetra-\atop\scriptstyle hedra\ \Delta}
C_{{1\over2}n(\bar S)}^{\alpha(\bar S\cap\partial\Delta)}
T[\alpha(\bar S\cap\partial\Delta)],
\label{eq:part2}
\end{eqnarray}
where $\bar S$ is a closed surface formed by connecting
the surfaces $S$ each  of which is defined inside a single tetrahedron,
$\alpha(\bar S\cap\partial\Delta)$ is a loop defined
by the intersection of $\bar S$ with the boundary (or faces)
$\partial\Delta$ of a tetrahedron, $n(\bar S)$ is the number of times
 $\bar S$ crosses a particular edge (related to $j$ by
$n=2j$), and $N:=2L$ is the cutoff of the values of $n$.

Eq.(\ref{eq:part1}) means that given a configuraton of
$j$'s on the tetrahedral
lattice, the set of all the possible surfaces $E(j)$
 inside a single tetrahedron is determined, and
each surface $S\in E(j)$ determines the coefficient
$C_j^{\alpha(S)}$. In this sense everything is dependent on the
given $j$'s and the partition function is the sum over
all the possible configurations of $j$'s.

Eq.(\ref{eq:part2}) means that given a closed surface
$\bar S$, the length $j$ of each edge is determined
by counting the number of times $\bar S$ crosses the edge, and
a loop around a tetrahedron is determined by taking
intersection of $\bar S$ and the faces of the tetrahedron.
In this sense everything is determined by the configuration
of a given closed surface, and the partition function
is the sum of all the possible configurations of surfaces on
the 3d lattice.

Here notice that the connection of two surfaces constructed in
respective tetrahedra which share a common face is not straightforward.
If one surface induces loop like (c) in Fig.~\ref{fig:spin}
and the other surface does like (d) in the same figure
on the common face, then the connected surface has
an irregular shape at the face;
otherwise, an equivalence class of surfaces can be well defined.
Therefore, the surface $\bar S$ is, in general, irregularly connected
in this sense.
In this paper we leave this problem open and consider
the irregularly connected ones as surfaces
unless otherwise explicitly stated.
We discuss this problem in Sec.~\ref{sec:discuss}.

The regulator $\Lambda(N)$ could be reconstructed
in the language of surfaces.
In this paper we disregard this issue and implicitly assume
the form Eq.(\ref{eq:lambda}).
In the following, we construct the coefficient $C_j^{\alpha(S)}$ and
the explicit partition function in terms of surfaces $\bar S$.

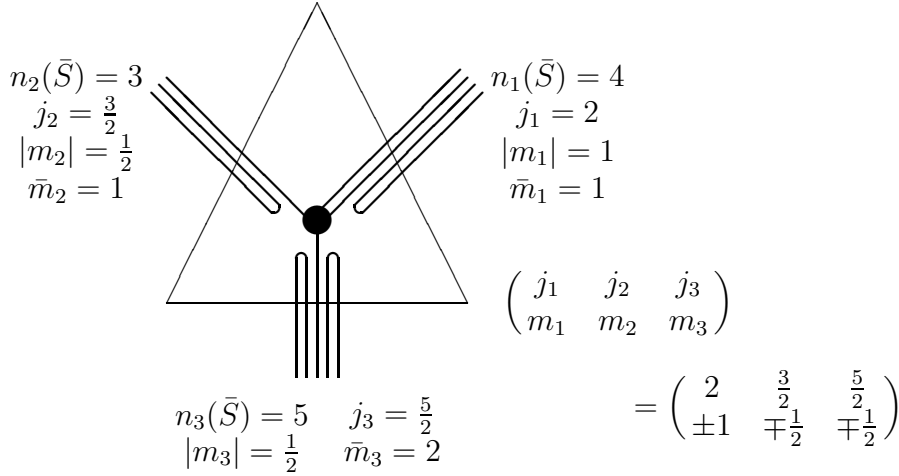
\begin{figure}
\unitlength=1.0mm
%\linethickness{0.6pt}
\begin{picture}(80.00,60.00)(-30.00,-20.00)
%\thicklines
\put(0.00,0.00){\line(1,0){40.00}}
\put(0.00,0.00){\line(1,2){20.0}}
\put(40.00,0.00){\line(-1,2){20.00}}
{\thicklines
\put(20.00,10.00){\line(-1,1){20.00}}
\put(15.00,13.00){\line(-1,1){16.00}}
\put(14.00,12.00){\line(-1,1){16.00}}
\put(14.00,13.00){\oval(2,2)[br]}

\put(20.00,12.00){\line(1,1){19.00}}
\put(20.00,10.00){\line(1,1){20.00}}
\put(25.00,13.00){\line(1,1){16.00}}
\put(26.00,12.00){\line(1,1){16.00}}
\put(26.00,13.00){\oval(2,2)[bl]}

\put(20.00,10.00){\line(0,-1){20.00}}
\put(17.20,6.00){\line(0,-1){16.00}}
\put(18.60,6.00){\line(0,-1){16.00}}
\put(21.40,6.00){\line(0,-1){16.00}}
\put(22.80,6.00){\line(0,-1){16.00}}
\put(17.90,6.00){\oval(1.4,1.4)[t]}
\put(22.10,6.00){\oval(1.4,1.4)[t]}

\put(20.00,11.00){\circle*{4}}
}
\put(52.00,25.0){\makebox(0,0)[cc]{$j_1=2$}}
\put(52.00,20.0){\makebox(0,0)[cc]{$|m_1|=1$}}
\put(52.00,15.0){\makebox(0,0)[cc]{$\bar m_1=1$}}
\put(52.00,30.0){\makebox(0,0)[cc]{$n_1(\bar S)=4$}}

\put(-12.00,25.0){\makebox(0,0)[cc]{$j_2={3\over2}$}}
\put(-12.00,20.0){\makebox(0,0)[cc]{$|m_2|={1\over2}$}}
\put(-12.00,15.0){\makebox(0,0)[cc]{$\bar m_2=1$}}
\put(-12.00,30.0){\makebox(0,0)[cc]{$n_2(\bar S)=3$}}

\put(30.00,-15.0){\makebox(0,0)[cc]{$j_3={5\over2}$}}
\put(10.00,-20.0){\makebox(0,0)[cc]{$|m_3|={1\over2}$}}
\put(30.00,-20.0){\makebox(0,0)[cc]{$\bar m_3=2$}}
\put(10.00,-15.0){\makebox(0,0)[cc]{$n_3(\bar S)=5$}}

\put(60.00,-0.0){\makebox(0,0)[cc]
{$\left(\matrix{j_1&j_2&j_3\cr m_1&m_2&m_3}\right)$}}
\put(80.00,-14.0){\makebox(0,0)[cc]
{$=\left(\matrix{2&{3\over2}&{5\over2}\cr
\pm 1&\mp{1\over2}&\mp{1\over2}}\right)$}}
\end{picture}
\caption{An example of the specific numbers for dual lines,
and the 3-j symbols determined by them.
All these numbers are determined by a (multiple) surface
on a fixed tetrahedral lattice.}
\label{fig:example}
\end{figure}

To construct $C_j^{\alpha(S)}$, we make one more observation
as follows.
In the third observation we made above,
we noticed the fact that since spin-$1\over2$ ``parallel transports"
placed on the same dual line are identical,
there are indistinguishable ways of contracting
the spin-$1\over2$ ``parallel transports" on a 2d dual lattice.
Let us be explicit about this statement.
Consider a component of the spin-$j$ matrix of an $SU(2)$ element
and write it as a sum of products of the components of
the spin-$1\over2$ matrix of the element.
For a general  $SU(2)$ element, it is
\begin{eqnarray}
&&D^{\ \ m}_{(j)\ m'}={1\over(2j)!}\sum_{n_1,n_1'=-1/2}^{1/2}\cdots
\sum_{n_{2j-1},n_{2j-1}'=-1/2}^{1/2}\nonumber\\
&&\times\sqrt{\prod_{\nu=1}^{2j-1}
\left[(j-{\nu\over2}+1)+2n_\nu(m-n_1-n_2\cdots-n_\nu)\right]}
\nonumber\\
&&\times\sqrt{\prod_{\nu=1}^{2j-1}
\left[(j-{\nu\over2}+1)+2n_\nu'(m'-n_1'-n_2'\cdots-n_\nu')\right]}
\nonumber\\
&&\times D^{\ \ n_1}_{(1/2)\ n_1'} D^{\ \ n_2}_{(1/2)\ n_2'}\cdots
D^{\ \ n_{2j-1}}_{(1/2)\ n_{2j-1}'}
D^{\ \ (m-n_1-n_2\cdots-n_{2j-1})}_{(1/2)\ (m'-n_1'-n_2'\cdots-n_{2j-1}')}.
\label{eq:element}
\end{eqnarray}
If $D^{\ \ m}_{(j)\ m'}$ is a component of
one of the matrices in Eq.(\ref{eq:mat}),
then it is non-vanishing only if $|m|=|m'|$,
and contains the same numbers of identical components of
the spin-$1\over2$ matrix in every term, the number depending on
the value of $|m|$.
(This fact is analogous to the case of the spin state of
a system consisting of identical spin-$1\over2$ particles.)
Because of this fact,
indistinguishable ways of doing the contractions
are present within each value of $|m|$.
Therefore, we recast Eq.(\ref{eq:6-j}) to write
\begin{eqnarray}
&&\left\{\matrix{j_1&j_2&j_3\cr j_4&j_5&j_6}\right\}=
\sum_{|m_1|,\cdots,|m_6|}(2-\delta_{\sum_{i}|m_i|,0})
\cos[\pi(j_3+j_6+m_1+m_4)]
\nonumber\\
&&\times
\left(\matrix{j_1&j_2&j_3\cr m_1&m_2&-m_3}\right)
\left(\matrix{j_4&j_5&j_3\cr m_4&m_5&m_3}\right)
%\nonumber\\
%&&\times
\left(\matrix{j_4&j_2&j_6\cr m_4&m_2&-m_6}\right)
\left(\matrix{j_1&j_5&j_6\cr m_1&m_5&m_6}\right),
\label{eq:s-term}
\end{eqnarray}
where the sums are taken over non-negative values and $m_i$ in
the summand takes the value $+|m_i|$ or the value $-|m_i|$,
the choice being determined by the requirement that
the 3-j symbols be non-vanishing.
Here, the alternative choice of the signs of all $m_i$'s makes
no difference.
Furthermore, if $D_{(j)}$ is one of matrices in Eq.(\ref{eq:mat}),
all terms in Eq.(\ref{eq:element}) for a fixed $|m|$ are identical.
Therefore, each term in Eq.(\ref{eq:s-term}) corresponds
to an equivalence class;
hence the values of $(|m_1|,|m_2|,\cdots,|m_6|)$
classify the loops on the dual lattice
for fixed values of $j$'s.
We can identify each 3-j symbol in Eq.(\ref{eq:s-term})
with one of the triangles on the 2d dual lattice
(see Fig.\ref{fig:latt})
with some configuration of parallel lines on the dual lines.
$2|m_i|$ is the number of parallel lines
(on the corresponding dual line with value $j_i$)
connected to  parallel lines on the other
dual lines at the triangle.
The number of disconnected parallel lines makes no contribution to
the value of the 3-j symbol.
An example of the relation between these numbers and a diagram
is partially given in Fig.\ref{fig:example}.

Therefore, the summand of Eq.(\ref{eq:CT}) is
\begin{eqnarray}
&&C_j^{\alpha(S)}T[\alpha(S)]=
(-1)^{\sum_ij_i}(2-\delta_{\sum_{i}|m_i|,0})
\cos[\pi(j_3+j_6+m_1+m_4)]
\nonumber\\
&&\times
\left(\matrix{j_1&j_2&j_3\cr m_1&m_2&-m_3}\right)
\left(\matrix{j_4&j_5&j_3\cr m_4&m_5&m_3}\right)
%\nonumber\\
%&&\times
\left(\matrix{j_4&j_2&j_6\cr m_4&m_2&-m_6}\right)
\left(\matrix{j_1&j_5&j_6\cr m_1&m_5&m_6}\right).
\label{eq:surf}
\end{eqnarray}
Correspondingly $E(j)$ is the set of all the possible values of
$(|m_1|,|m_2|,\cdots,|m_6|)$ for fixed $j$'s.
Now, Eq.(\ref{eq:part1}) has been explicitly constructed.

The next task is to express the coefficient
entirely in terms of surfaces $\bar S$ without
$j$'s, as in Eq.(\ref{eq:part2}).
Here we have a problem due to the irregularly connected surfaces
as noticed above.
We disregard this problem and assume that surfaces
we are considering are all regularly connected ones.

We introduce a set of non-negative half-integers $|m|$'s and
non-negative integers $\bar m$'s to specify surfaces,
where a pair $(|m|,\bar m)$ is assigned to each edge.
In other words, given a surface $\bar S$
on a fixed tetrahedral lattice, $(|m|,\bar m)$ for an edge is
determined as follows.
Consider a face to which the edge belongs.
The intersection of $\bar S$ and the face is seen as a portion of
a loop or parallel lines on the dual lines at the face.
The number of parallel lines
(on the dual line corresponding to the edge)
connected to parallel lines on the other dual lines
at the face is $2|m|$ and
the number of disconnected parallel lines on the same dual line
is $2\bar m$.
$j$ is defined by $j:=|m|+\bar m$.
(Accordingly $n(\bar S)=2(|m|+\bar m)$.)
A 3-j symbol can be assigned to the face with the values $j$ and
$m=+|m|$ or $-|m|$ such that the symbol is non-vanishing.
An example of the specification of these numbers is shown
in Fig.\ref{fig:example}.
Therefore, combining the four faces of a tetrahedron,
$C_j^\alpha(S)T[\alpha(S)]$ for the tetrahedron can be determined
by Eq.(\ref{eq:surf}).
The substitution into
$S_{{1\over2}n(\bar S)}^{\alpha(\bar S\cap\partial\Delta)}
T[\alpha(\bar S\cap\partial\Delta)]$
in Eq.(\ref{eq:part2})
completes our construction of the partition function
 in terms of surfaces $\bar S$.
The sum is taken over all the possible values of
all $|m|$'s and $\bar m$'s, which classify the
surfaces $\bar S$.

\section{Discussion}
\label{sec:discuss}

In the previous section we described how the Ponzano-Regge model
can be reformulated as a theory of surfaces.
For a given 3d simplex lattice, we constructed a set of surfaces
and presented a way that the set of surfaces determines the lengths
of all the edges on the simplexes.
A key point is that a geometry (a set of the lengths of the edges)
is not given on the simplex lattice
but is determined by the set of surfaces.
Therefore, the main geometrical object is not the simplex lattice
but the set of surfaces.
We can therefore view the simplex lattice just as a tool
to compute the value of
the partition function.

This situation is analogous to problems of 2d Euclidean geometry.
One can study geometry with geometrical
objects themselves (i.e. lines, circles, triangles, etc).
Instead, one can introduce a set of coordinates and
convert a geometrical problem to an algebraic problem.

In our case, a set of coordinates, which is a tool for converting
a geometrical problem to an algebraic problem,
corresponds to a simplex lattice and the geometrical object is
a set of surfaces.
Therefore, our formulation here emphasizes that
our theory of surfaces is a geometrical model and
have geometrical information by itself
without referring to a simplex lattice.
However, in order to compute concrete quantities like the values
of lengths and the partition function,
it is convenient to use a simplex lattice.
In order to show this fact from our formulation
we describe a possible scenario as follows.

Given a set of continuous surfaces in a 3d space-time
manifold (not a simplex manifold), triangulate the space-time.
Note that without any geometrical information the triangulation
does not have the notions of lengths, areas, lines, triangles, etc.
The tetrahedra are just regions separated from one another.
The faces are just surfaces and the edges are just curves
in the space-time.
The triangulation simply specifies the places of the edges
and the faces of the tetrahedra in the space-time.

Then consider the triangulation as a simplex lattice and
replace the continuous surfaces by discretized surfaces
constructed from elementary surfaces.
This replacement can be done uniquely
once a simplex lattice is defined.
Then according to the prescription described
in the Sec.~\ref{sec:surface},
the set of surfaces determines the values of quantities we need.
By making variations of surfaces on the lattice fixed,
we can compute the partition function in principle.
Since the partition function does not depend on
the choice of the triangulation, our theory of surfaces
is geometric in the sense described above.
This scenario could be also considered
as a regularization of surfaces.

The formulation we developed has the following features:

\noindent
(1) The relation to the loop representation of canonical quantum
general relativity in 3-dimensions is transparent.
That is, the intersections of the closed surfaces in $(2+1)$
space-time with any 2-space are loops in canonical theory.

\noindent
(2) Since this formulation relies on geometric objects,
namely surfaces, the extension into 4-dimensions could be possible,
in contrast to the 6-j symbolic formalism of the Ponzano-Regge
model. In short the idea  is the following.
Since the intrinsic curvature in the $n$-dimensional Regge
calculus is concentrated in $(n-2)$-simplexes, the surfaces,
which are formed by faces dual to the $(n-2)$- simplexes,
could play an important geometrical role in any $n$-dimensions.

\noindent
(3) This formulation is purely a theory of surfaces.
In other words, since it does not include any field variables,
it is coordinate-free on the surface and
background-free in spacetime.
Thus, this formulation could be seen as a discretization of
a genuinely background independent string theory.

The continuation of this work is to reconstruct the regulator
$\Lambda(N)$ in the language of surfaces,
which has been disregarded in this paper.
The next step is to reconsider
within this formulation
the problems that have already been solved in 3d quantum gravity
in a way
that one could perhaps develop technologies and insights
that could be extended into 4d quantum gravity.
In particular, the problem of the
construction of the inner product of canonical theory
in the loop representation could be attacked along these lines,
as pointed out by Rovelli \cite{rovelli}.

The problems of

\noindent
(i) the irregularly connected surfaces noticed in Sec.~\ref{sec:surface},

\noindent
(ii) the regularization of infinities in the partition function,

\noindent
(iii) the use of equivalence classes of surfaces defined
in Sec.~\ref{sec:surface},

\noindent
(iv) the derivation of homotopy classes, which label the loop states,
and

\noindent
(v) the rederivation of the inner product of
the (2+1) loop representation

\noindent
might be closely related.
For this formulation
to be equivalent to the loop representation,
in which loop states are labelled by homotopy classes,
the irregularly connected surfaces might be irrelevant
to physical degrees of freedom and might be eliminated
from the formulation.
This possibility comes from the fact that
folded parts of surfaces produce retracing segments of loops
on a 2-space, which are known to make no contribution in the theory,
and the elimination of the folded parts results in the elimination
of irregularly connected surfaces.
This elimination of the irregularly connected surfaces might
make the equivalence classes of surfaces relate to the homotopy
classes.
Furthermore, this elimination of unnecessary degrees of freedom
might lead to the regularization of infinities in the language
of surfaces.
Therefore, the rederivation of the inner product of
the loop representation from this formulation might be tractable.

In addition to the issue of the inner product,
the question of how classical spacetime
geometry can be recovered from the purely topological
language of surfaces
(or loops in the case of canonical theory)
could be studied.\footnote{A different strategy to recover
a semiclassical flat geometry from the loop-space formulation
was applied in Ref. \cite{mapM,vacuum}.}
Based on the fact that in the large $j$ limit the Regge action
must be recovered in the exponent of the summand, we expect that
in the limit the coefficient $C_j^{\alpha(S)}$ gives
the exponential of the sum  of the
deficit angles over a (multiple) surface.
Each deficit angle is equal to the sum of $\theta_i$
around the corresponding edge.
Another line of work we could consider is
to regard this formulation
as a string model in 3d or 4d spacetime (which, in the 6-j
symbolic formalism, has been
shown to be  equivalent to quantum general relativity
in case of 3-dimensions \cite{ooguri}).
The ``duality" between the loop representation and the string theory
was suggested by Baez.\cite{baez}
Our formulation could make the idea concrete.
Work is in progress along these lines.

The present work emerged based on an observation of
the 6-j symbols, building blocks of the Ponzano-Regge model.
Some formulations
(i.e. Boulatov model\cite{boulatov}, spin-network\cite{spin-net})
are known to be related to the Ponzano-Regge model.
However, relations of the present work to these models
are not well understood at present.
Certainly it is interesting to investigate them.

\vskip0.3cm\noindent
The author thanks Carlo Rovelli for discussions and comments.

\end{document}